\documentclass[twocolumn,showpacs,prl,superscriptaddress,amsmath,amssymb]{revtex4-1}

\usepackage{graphicx}
\usepackage{color}

\usepackage{graphicx}
\usepackage{dcolumn}
\usepackage{bm}
\usepackage{times}
\begin{document}

\title{Tuning the disorder in  superglasses}

\author{Derek Larson}
\affiliation{Department of Physics,\\
National Taiwan University, No. 1, Sec. 4, Roosevelt Rd., Taipei 106, Taiwan}
\author{Ying-Jer Kao }
\email{yjkao@phys.ntu.edu.tw}
\affiliation{Department of Physics,\\
National Taiwan University, No. 1, Sec. 4, Roosevelt Rd., Taipei 106, Taiwan}
\affiliation{	Center for  Advanced Study in Theoretical Science,   \\
National Taiwan University, No. 1, Sec. 4, Roosevelt Rd., Taipei 106, Taiwan}

\date{\today}

\begin{abstract}

We study the interplay of superfluidity, glassy and magnetic orders in the XXZ model with random  Ising interactions on a three dimensional cubic lattice.
In the classical limit, this model reduces to a $\pm J$ Edwards-Anderson Ising model with concentration $p$ of ferromagnetic bonds, which hosts a glassy-ferromagnetic transition at  a critical concentration $p_c^{\rm cl}\sim 0.77$. Our quantum Monte Carlo simulation results  show that quantum fluctuations stabilize the coexistence of superfluidity and glassy order ( ``superglass''), and shift the (super)glassy-ferromagnetic transition to $p_c> p_c^{\rm cl}$. In contrast, antiferromagnetic order coexists with superfluidity to form a supersolid, and the transition to the glassy phase occurs at a higher  $p$. 

\end{abstract}

\pacs{75.50.Lk, 75.40.Mg, 05.50.+q}
\maketitle
\paragraph*{Introduction--}
Spin glasses  are frustrated magnetic systems with quenched disorder,  hosting glassy phases with  extremely slow dynamics\cite{binder:86}. 
Traditionally, the simplest model that exhibits  spin glass (SG) behavior is the $\pm J$ Ising model, or Edwards-Anderson (EA) model\cite{edwards:75}, in which Ising spins interact via randomly distributed nearest-neighbor (NN) ferromagnetic (FM) and antiferromagnetic (AFM)  bonds  with probability $p$ and $1-p$ respectively. This model has been studied extensively with Monte Carlo simulations and exhibits a finite temperature glass transition in three dimensions (3D)~\cite{kawashima:96, Ballesteros:2000kx}.
It has been shown that this model has a second order transition from a SG to a FM (AFM) phase at a critical concentration $p_c=p_c^{\rm cl}\sim 0.77\,(p_c=1-p_c^{\rm cl}\sim 0.23) $ as one increases (decreases) the fraction of FM bonds, and there exists reentrance of the spin glass phase in the temperature-disorder phase diagram\cite{hartmann:99b, hasenbusch:08b,Ceccarelli:2011kx}.
One natural question to ask is how  the transition can be changed by quantum fluctuations~\cite{Markland:2011kx, tam:10, Yu:2012vn}.  In particular, understanding the behavior of  granular superfluidity in a frozen amorphous structure may give us hints on the microscopic mechanism of the formation of supersolids (SS)~\cite{Kim04Nature,*Kim:2004fk}. 

 The observation of  supersolidity in  solid Helium 4 (He$^4$) \cite{Kim04Nature,*Kim:2004fk} has spurred immense interests on the connection between superfluidity and disorder in bosonic systems. Strong experimental evidence suggest that disorder may play a role in how the supersolid forms\cite{Rittner:2007fk,*Rittner:2006uq}.   A recent experiment observed ultraslow dynamics in solid He$^4$\cite{hunt:09}, suggesting a glassy type of supersolid,  or a ``superglass'' (SuG) ~\cite{brioli:08,wu:08,tam:10,Yu:2012vn}. Quantum Monte Carlo (QMC) studies on the extended hardcore Bose-Hubbard model with random frustrating interactions on a 3D cubic lattice
 indicate that glassiness can coexist with superfluidity, and quantum fluctuations and random frustration are both crucial in stabilizing the superglass state\cite{tam:10}. 

While the  phase transitions into superfluidity or quantum solid/glassy states have
been studied in some detail,  the nature of the various transitions into
a SS/SuG are yet left unexplored.  In this {\it Letter}, we study  the XXZ model with random Ising interactions, which maps to an extended hardcore Bose-Hubbard model with random frustrating interactions,  in order to
characterize  transitions by tuning the amount of disorder present in 
the interactions.
We demonstrate that the presence of quantum exchange terms
greatly increases the temperature dependence of the SuG-FM transition by pushing the low temperature phase boundary  to a higher critical concentration than $p_c^{\rm cl}$. The 
SuG-SS phase boundary is also drawn to a  higher critical concentration, and  an asymmetry for SuG-SS and SuG-FM transitions arises.

\paragraph*{Model--}

The Hamiltonian for hardcore boson with a random NN interaction is
\begin{equation}
{\mathcal H} = - \sum_{\langle i,j\rangle} V_{ij} (n_i - 1/2)(n_j - 1/2) - t \sum_{\langle i,j\rangle} (b_i^\dagger b_j + b_i b_j^\dagger) ,
\end{equation}
where $\langle i,j \rangle$ indicates the NN lattice sites.  
$n_i $  is the number operator for hard-core bosons 
at lattice site $i$,
$t$ is the hopping parameter and $V_{ij}$ are interactions with a bimodal distribution
given as 
\begin{equation}
p(V_{ij}) = p \delta (V_{ij} - V) + (1 - p) \delta (V_{ij} + V) .
\label{eq:prob}
\end{equation}
By a transformation of the bosonic operators  to spin-1/2 operators: $S^z_i=n_i -1/2, S_i^-=b_i$, and $S_i^+= b^\dagger_i$,
this model can be readily mapped into the standard XXZ model:
\begin{equation}
{\mathcal H} = - \sum_{\langle i,j\rangle} J_z S_i^z S_j^z - \frac{1}{2}J_{xy}\sum_{\langle i,j\rangle} (S_i^- S_j^+ + S_i^+ S_j^-),
\label{eq:hamiltonian}
\end{equation}
where $J_{xy}=2t$ and $J_z=V_{ij}$. This model reduces to the classical EA model if $J_{xy}=0$. The FM and AFM phases in the spin model correspond to  quantum solid phases with $(0,0,0)$ and $(\pi,\pi,\pi)$ ordering vectors, respectively. In the following, we will use the spin language for simplicity.

There exist both diagonal and off-diagonal long-range orders in this model. Possible diagonal long-range orders are FM, AFM and SG, with corresponding order parameters:  magnetization $m= \frac{1}{N}\sum_i S_i^z$ (FM), staggered magnetization  $m_s = \frac{1}{N}\sum_i (-1)^i S_i^z$ (AFM), and the
Edwards-Anderson order parameter (SG), defined as
\begin{equation}
q_{\rm EA} =\frac{1}{N} \left[{ \sum_i \langle S_i^z \rangle^2 }\right]_{\rm av},
\end{equation}
where $\langle \cdots \rangle$ denotes a thermal average and
$[\cdots]_{\rm av}$ an average over disorder realizations. As the
EA order parameter will also capture FM and AFM ordering, one must look for a non-zero $q_{\rm EA}$
while the other order parameters remain zero in order to identify an SG phase. 
 
To determine the phase transition point, we look at the Binder cumulants \cite{binder:81b} for the order parameter, which should cross at the transition point for different system sizes. 
However, the existence of corrections to scaling cause pairs of small sizes to intersect away from the true critical point.  To improve the accuracy of the measurement, we look at the series of intersection points created by successive pairs of sizes $L$ and $L + 1$ which should converge to the exact value as a power law with the exponent determined by the leading correction to scaling.  To get statistical estimates of these crossing points, we perform a bootstrap resampling on the raw data ($N_\text{boot} = 100$), fitting a 2nd-order polynomial to each size, and calculating the resulting intersection.
The Binder cumulants of the magnetization  and staggered magnetization   are
\begin{equation}
g_{\rm m} = \frac{1}{2} \left(3 - \frac{\left[ \langle m^4 \rangle\right]_{\rm av}}{\left[\langle m^2 \rangle\right]_{\rm av}^2} \right);\quad g_{\rm sm} = \frac{1}{2} \left(3-\frac{\left[\langle m_s^4 \rangle\right]_{\rm av}}{\left[ \langle m_s^2 \rangle\right]_{\rm av}^2}\right)
\label{eq:g_m}
\end{equation}
which approach one in the FM/AFM phase and goes to zero otherwise. 
We measure the superfluid density  through winding
number fluctuations\cite{Pollock87PRB}, 
\begin{equation}
\rho_s = \frac{1}{3 \beta N t} \left[\sum_{k = x,y,z} \langle W_k^2 \rangle \right]_{\rm av},
\label{eq:rho_s}
\end{equation}
where $W_k$ is the winding number
along the $k$ direction.
In our simulation, we identify the SS phase by the coexistence of AFM and SF orders, and the SuG phase by the coexistence of SG and SF orders. 
\begin{figure}[tbp]
\includegraphics[width=1\columnwidth]{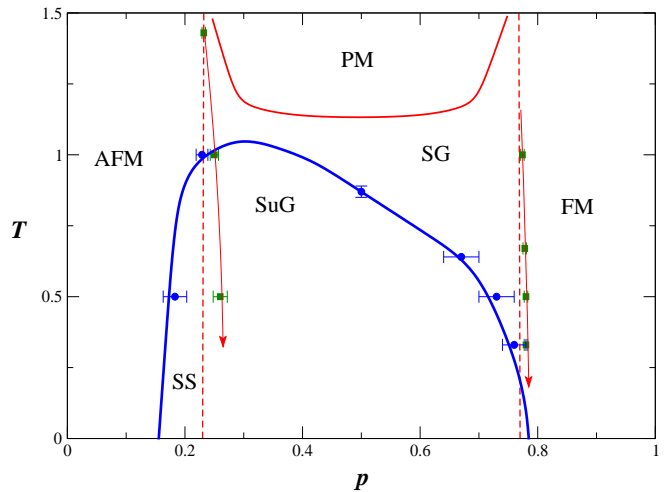}
\vspace*{-0.6cm}
\caption{(Color online)
$T$ vs $p$ phase diagram.  The dashed lines indicate the position
of the classical transitions SG-(A)FM, and the solid lines with arrows 
show the trend of the phase boundary shifts we find in the quantum model. The blue
circles are our estimates for the superfluid transition, with the approximate phase
boundary drawn as the blue arc. Lastly, the red arc denotes the PM-SG phase boundary.  
}
\label{fig:phase}
\end{figure}

\paragraph*{Method--}

We use the Worm Algorithm QMC~\cite{Prokofev:1998kx, *Prokofev:1998vn} to study
this model, as well as the Stochastic Series Expansion (SSE)~\cite{sandvik:99} with the parallel tempering~\cite{hukushima:96,*Melko:2007fk} near the FM boundary to access low temperatures\cite{SM}.
We simulate many different realizations of the disorder--sets of interactions $V_{ij}$--for
a given parameter set.  In the following, we choose $V/t=4$ in order to maximize the extent
of the superglass phase at $p=0.5$\cite{tam:10}.  Each realization is simulated independently
and equilibrium is determined by reaching measurements that stabilize within our error bars. 
Table ~\ref{tab:simparams} contains the details of our simulation parameters.
For our worm algorithm implementation a Monte Carlo (MC) step is defined as 
$N_{\rm sites}=L^3$ completed worm loop updates, while for SSE we define it the same 
as Ref.~\cite{sandvik:99} with the addition of one replica exchange sweep.
Equilibration times vary across phases: worm update loops tend to be long while   
in SF phases, and under increasing ferromagnetic order the worms have a low probability of hopping
producing short loops. This effect is significantly reduced in our SSE implementations since 
each MC step has an adaptively determined number of operator loop updates.

\begin{figure*}[tbp]
\begin{center}
\includegraphics[width=1.9\columnwidth]{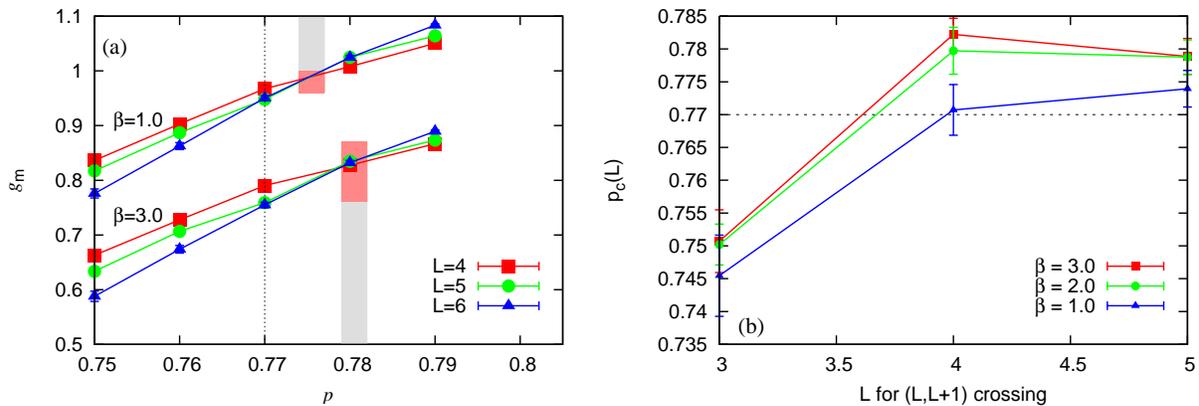}
\end{center}
\vspace*{-0.5cm}
\caption{(Color online)
(a) Finite size scaling of $g_m$ at $\beta = 1$ and 3
to determine the SG-FM transition. Data for $\beta = 1$ are shifted vertically for clarity. Shaded areas indicate the crossing for different sizes. 
Dotted line shows the position of the zero-temperature classical transition ($p_c^\text{cl}$).  Our $p_c$ estimates show stronger reentrance.
(b) Results of bootstrap estimates for the crossing points for adjacent sizes. Dotted line shows the position of the zero-temperature classical transition. Deviation from the classical transition point is clearly seen in the large size data points at $\beta=2$ and 3. 
}
\label{fig:FM}
\end{figure*}

\begin{figure}[tbp]
\includegraphics[width=1\columnwidth]{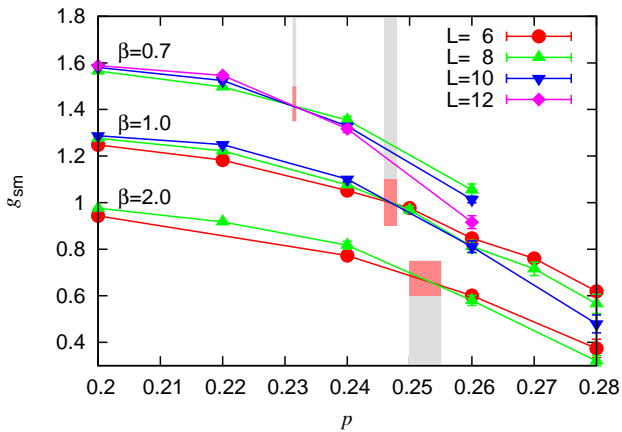}
\vspace*{-0.0cm}
\caption{(Color online)
Finite size scaling of the staggered magnetization cumulant $g_{sm}$. Data for $\beta = 1$ and $\beta = 0.7$
are shifted vertically for clarity.  Results at $\beta = 0.7$ are in line with classical behavior while  
$\beta = 1,2$ suggest the SuG-SS transition has been drawn {\it into} the classical
SG region. 
}
\label{fig:AFM}
\end{figure}

\begin{table}[!tb]
\caption{
Parameters of the simulations
\label{tab:simparams}}
\begin{tabular*}{\columnwidth}{@{\extracolsep{\fill}} l l r r r r l}
\hline
\hline
&$\beta$ & $L$ & $p$ range & $\sim N_\text{samp}$ & $\sim N_\text{sweep}$ 
\\
\hline
&0.70 &    6 & 0.20 - 0.26 &   600 &     8000 & \\
&0.70 &    8 & 0.20 - 0.26 &   400 &    16000 & \\
&0.70 &   10 & 0.20 - 0.26 &   500 &    64000 & \\
&0.70 &   12 & 0.20 - 0.26 &   250 &   128000 & \\[2mm]

&1.00 &    3 & 0.70 - 0.80 &  2000 &     4000 & \\
&1.00 &    4 & 0.75 - 0.78 &  2000 &    16000 & \\
&1.00 &    5 & 0.75 - 0.78 &   800 &   128000 & \\
&1.00 &    6 & 0.75 - 0.78 &   150 &   256000 & \\
&1.00 &    6 & 0.12 - 0.28 &  1000 &    16000 & \\
&1.00 &    8 & 0.12 - 0.28 &   400 &    16000 & \\
&1.00 &   10 & 0.12 - 0.28 &   400 &    64000 & \\[2mm]

&2.00 &    6 & 0.20 - 0.28 &   500 &     8000 & \\
&2.00 &    8 & 0.20 - 0.28 &   500 &   128000 & \\[2mm]

&SSE\footnote{Various temperatures are simulated simultaneously in the parallel tempering. }  &    3 & 0.75 - 0.79 &  4700 &    16000 & \\
&SSE  &    4 & 0.75 - 0.79 &  4000 &    16000 & \\
&SSE &    5 & 0.75 - 0.79 &  4000 &    16000 & \\
&SSE  &    6 & 0.75 - 0.79 &  4000 &    32000 & \\

\hline
\hline
\end{tabular*}
\end{table}

\paragraph*{Results--}

Figure~\ref{fig:phase} summarizes our QMC results in a temperature-disorder ($T$-$p$) phase diagram. At high temperatures, we expect the model behaves classically, and is described by the 3D EA model. The classical phase diagram is symmetric about $p=0.5$ if one identifies  the FM with the AFM regime.  The classical 3D EA model undergoes a $T=0$ SG-FM transition at $p_c^{\rm cl} \sim 0.770$, and shows slight reentrant behavior at finite $T$ (red dashed lines). It is suggested that the finite-temperature SG-FM transition belongs to a new universality class with  critical exponents $\nu = 0.96(2)$ and $\eta = -0.39(2)$~\cite{Ceccarelli:2011kx}. The SG-FM,  PM-SG and  PM-FM transition lines meet at a multi-critical point $p=p^*$ (not shown), which sits on the Nishimori line\cite{nishimori:81,Ceccarelli:2011kx}, and the PM-SG transition for $1-p^*< p <p^*$  belongs to the Ising  spin glass (ISG) universality class\cite{Hasenbusch:2007ys}.   

Entering the quantum regime,  the SG-FM transition 
appears to become more temperature dependent and deviate from the classical behavior.
Likely this is due to the encroaching superfluid order and subsequent competition.
Meanwhile, the AFM state allows for superfluidity, creating a region of supersolidity within
the arm of the superfluid transition line, which may be why the AFM state is more
stable in the quantum model.  Furthermore, the SuG-SS line pushes deeper into the 
SG phase,  suggesting interesting correlation between the AFM and SF orders. 

The results for our analysis of the magnetic ordering are presented in Fig.~\ref{fig:FM}.  
Already at $\beta = 1$ we see evidence for a shift in the phase boundary.
By $\beta = 3$ in Fig.~\ref{fig:FM} $(a)$,
it is clear there would not be a transition for $p < 0.78$, given the position of the data
crossings as shown.
This is markedly different than the classical results.  The classical transition line $p_c^{\rm cl}$ 
as a function of $\beta$ is only weakly temperature dependent, shifting less than
a percent between the multicritical point and zero temperature\cite{Ceccarelli:2011kx}.  Here,
that shift is at least 4 times as large. 
We have studied the superfluid transition as well for these parameters, and our estimates 
(see Fig.~\ref{fig:phase}) are only rough due to the non-trivial nature of the superfluid
scaling form.  For a 3D XY model, one expects $\rho_s\sim~1/L^\alpha$ with $\alpha = 1$
away from quantum criticality.  However, we find $\alpha>1$ as the model enters the glass
phase with, for example, $\alpha \approx 1.4-1.8$ for the transition at $p=0.5$~\cite{larson:12b}.  
 
The specific mechanism behind the increased temperature dependence of the SuG-FM
transition should be rooted in how superfluidity
favors the spin glass phase over the ferromagnetic phase.    
One can imagine that  clusters of FM-ordered spins are suppressed as the system can gain kinetic energy by destroying local ferromagnetic order and allowing particles to hop. On the  other hand, glassy clusters more readily allow superfluid fluctuations
amongst them, perhaps following lines of sites that are weakly constrained
due to frustration.  Thus  a broader
spin glass (thus superglass) phase is stabilized and  the transition line is moved into the FM phase.  
 
\begin{figure}[tbp]
\includegraphics[width=0.8\columnwidth]{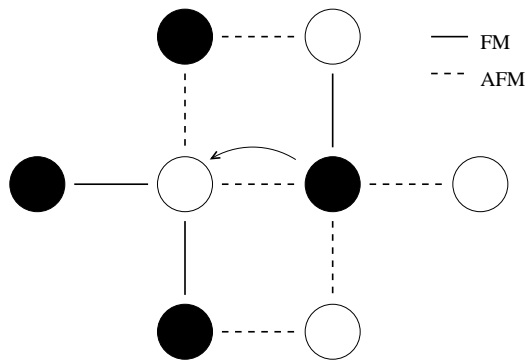}
\vspace*{-0.0cm}
\caption{
 A typical 2D snapshot near the SuG-SS line.  
The hop move indicated is energetically neutral with respect to the bonds  and
destroys the local AFM order.  However, the new state allows fewer fluctuations
so is not favored with respect to entropy. 
}
\label{fig:hop}
\end{figure}

We also looked at the behavior on the other side, $p < 0.5$, which is different 
due to the asymmetry of the ground state against quantum fluctuations.  Notably, we find
superfluid transitions at higher temperatures:  $\beta_c (1-p) < \beta_c (p)$.
Figure~\ref{fig:AFM} shows data for the Binder cumulant using the
staggered magnetization, $g_{sm}$ at $\beta = 1$ and $2$. 
The staggered magnetization data suffer from larger FSE and subsequently require
larger system sizes to achieve comparable precision, restricting the current study
from exploring lower temperatures. 

The position of the crossing at $\beta = 1$ lies around $p_c \sim 0.25$ indicating that
the SuG-SS line is noticeably shifted from classical behavior.  The data at lower $T$ 
are less conclusive due to larger sizes being out of reach at present, but suggest an
even larger shift.   Above the SF transition at $\beta=0.7 (T=1.43)$, the AFM-SG transition agrees with the classical transition point $p=1-p_c^{\rm cl}=0.23$. Thus, the stronger the quantum mechanical nature becomes--the
deeper into the superfluid behavior we go--the stronger the AFM order appears. 
This is at first counterintuitive, but we can provide a simple, consistent picture for both this  and
the SuG-FM behavior.  When an FM bond is satisfied, i.e. two neighboring sites are in the
same state, this forbids hopping along that bond.  Conversely, when an AFM bond is satisfied,
the two sites are in an opposite state and will allow the system to gain kinetic energy through
hops.  As this is energetically favorable, we can consider the AFM bonds to have a larger
\textit{effective} bond strength.  Consequently, the fraction of frustrating FM bonds it
would require to destroy the AFM order should rise, and we see the present shift in phase
boundary.

Speaking more generally, we can consider the change in the entropy of the system,
with respect to quantum fluctuations, when moving towards higher AFM or FM order.    
Under pure FM order no hopping is allowed, while under pure AFM order each site has
the ability to engage in virtual hopping.  
For the illustration purpose, Fig.~\ref{fig:hop} shows a typical local 2D snapshot of a system
with AFM order being frustrated by FM bonds.  The indicated move would go against three
NN bonds while satisfying three others, making the diagonal component energetically
neutral, and  this would turn the local AFM order into FM order.  However, after this move
the system is now considerably more constrained regarding the number of hoppings 
available.  Originally, the two interior sites could be involved with seven different
hop moves, but these sites now have a single move: to return back to the original
state. In this way, the AFM state is favored as it allows for more gain in
kinetic energy.

\paragraph*{Summary--}

We have presented our QMC results of a model exhibiting
a superglass phase in order to study the phase transitions
achieved by directly tuning the level of disorder present.
These results indicate that the addition of the exchange terms act
to stabilize the classical spin glass phase against the formation of
ferromagnetic clusters which impede hopping. In addition, the favoring
of AFM bonds to FM bonds leads to a shift in the SuG-SS phase boundary. 
An interesting issue not addressed
here is whether there is always a SG phase between the FM and SuG
phases.  Recent work\cite{Pollet:2009fk} on disordered Bose systems suggests that
this may be true.  Unfortunately, our current precision is not high enough to be conclusive.
Future work is required to clarify this issue.  

\begin{acknowledgments}

We thank Anders Sandvik and Markus M{\"u}ller for helpful discussions, and P. C. Chen for the use of his cluster at NTHU.
We are grateful to National Center for High-Performance Computing, and Computer and Information Networking Center at NTU for the support of high-performance computing facilities.
This work was partly supported by the NSC in Taiwan through Grants No. 100-2112-M-002 -013 -MY3, 100-2120-M-002-00 (Y.J.K.), and by NTU
Grant numbers  10R80909-4 (Y.J.K.). Travel support from  National Center for Theoretical Sciences is also acknowledged.

\end{acknowledgments}

\section{
Supplementary Material}

\begin{figure*}[b]
\includegraphics[width=0.9\columnwidth,clip]{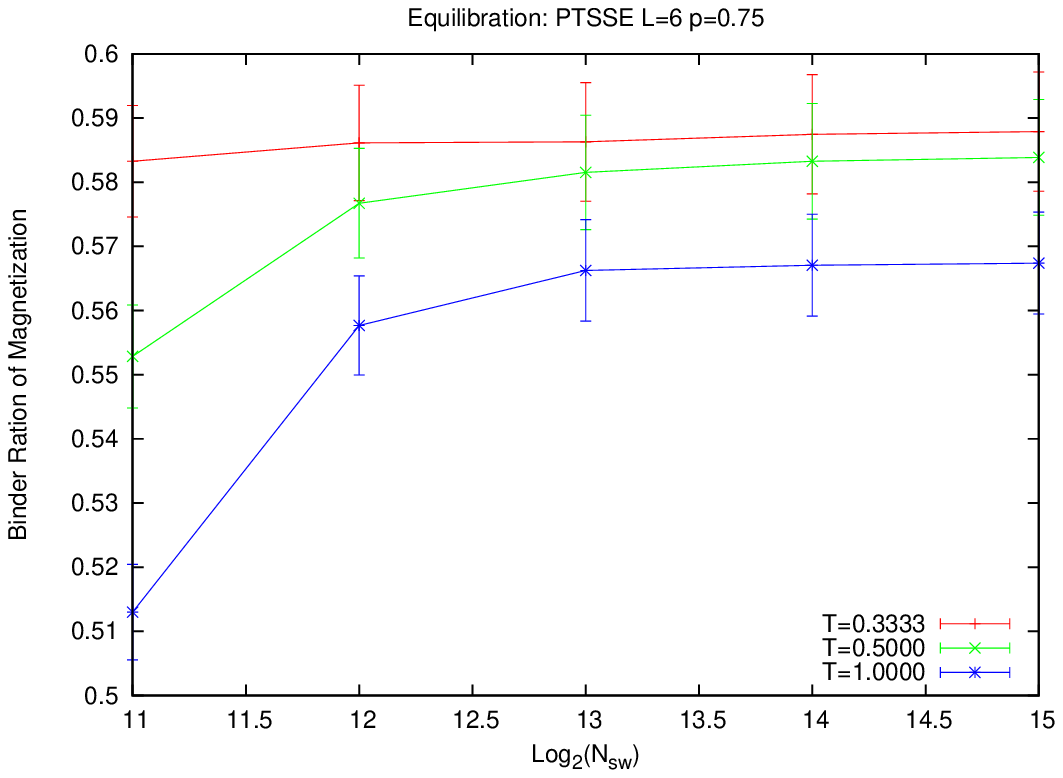}
\includegraphics[width=0.9\columnwidth,clip]{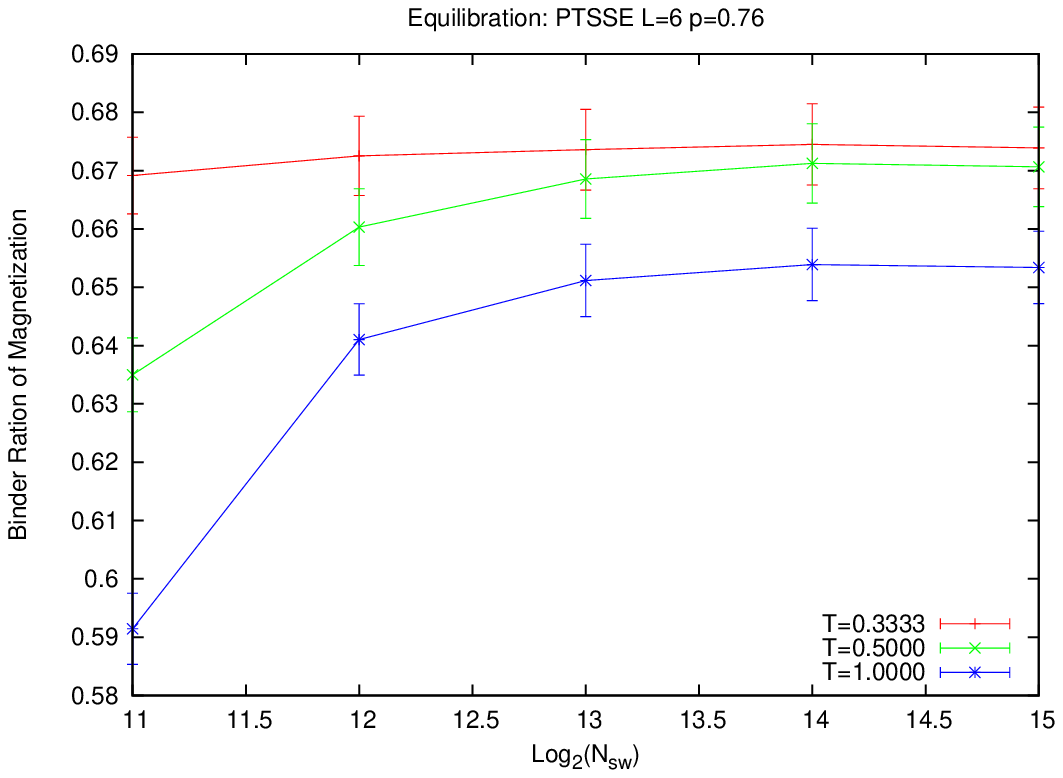}
\includegraphics[width=0.9\columnwidth,clip]{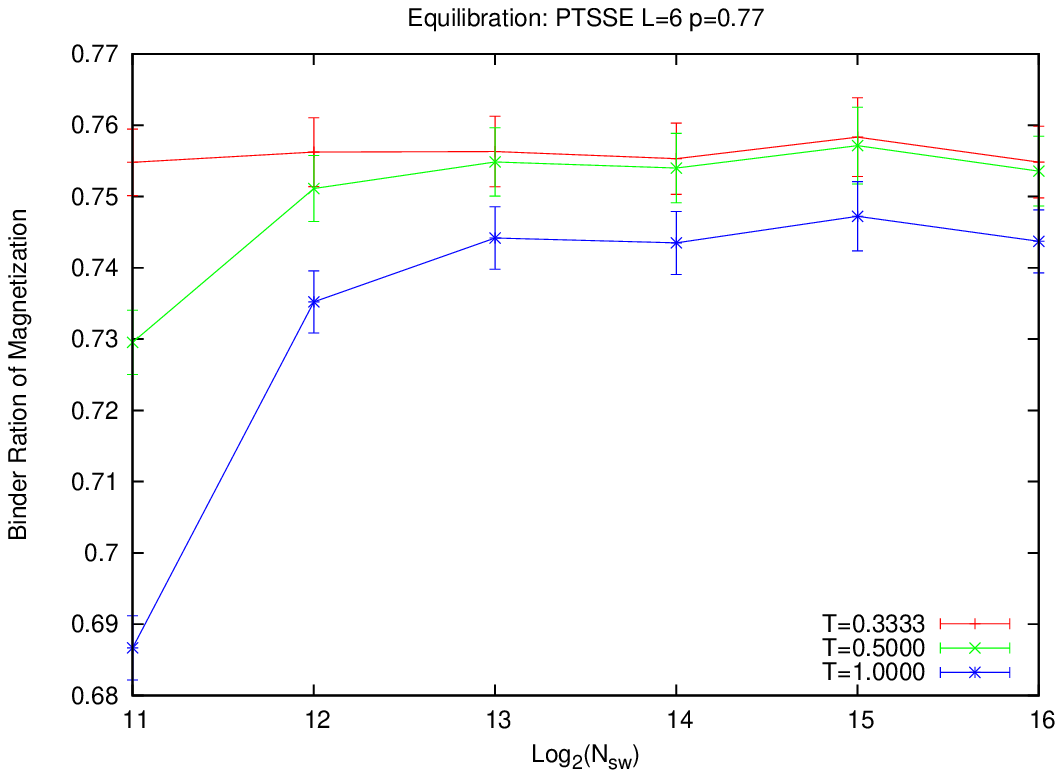}
\includegraphics[width=0.9\columnwidth,clip]{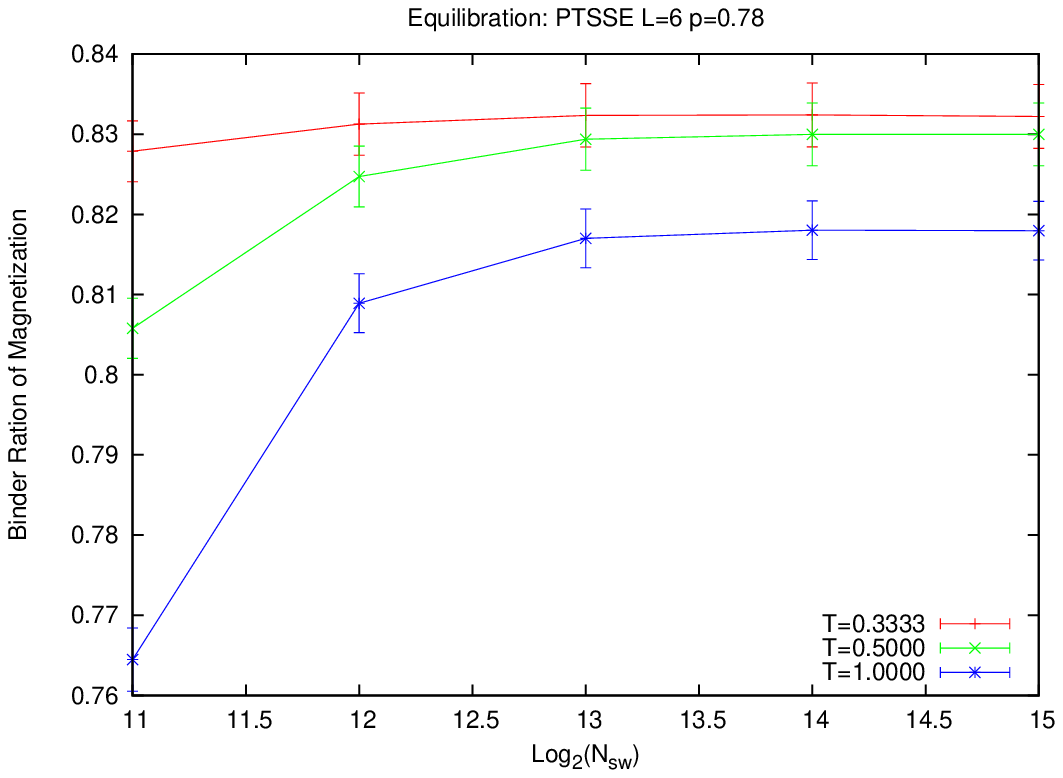}
\includegraphics[width=0.9\columnwidth,clip]{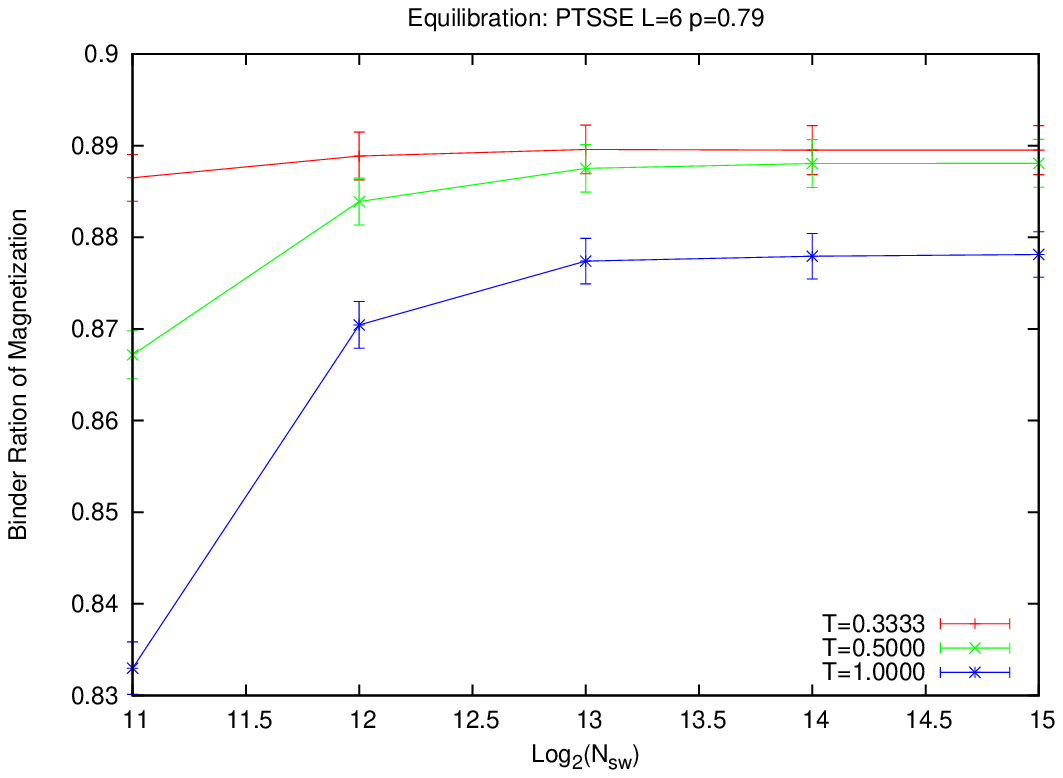}
\vspace*{-0.0cm}
\caption{ (Color online)
Equilibration data for the SSE at the largest system size $(L=6)$ for $p=0.75 - 0.79$. Smaller sizes are all more strongly equilibrated.
}
\label{fig:EQ}
\end{figure*}

Figure~\ref{fig:EQ} shows the equilibration data of the magnetic Binder cumulant, using the  stochastic series expansion (SSE) quantum Monte Carlo with parallel tempering (PT), near the ferromagnetic transition for $L=6$. Each data point is obtained by first equilibrating for a given number of sweeps ($N_{\rm sw})$, and the perform measurements for the same number of sweeps. For example,  the point at $N_\text{sw}=2^{10}$ will have1024 sweeps for equilibration and then 1024 sweeps of measurement.   In addition, averaging over different disorder realizations is carried out. It is clear that the simulations are well-equilibrated in the doping range of interests, $p=0.75 - 0.79$, and at all temperatures $T=0.3333,0.5000,1.0000$. Smaller sizes are all strongly equilibrated.  

\bibliography{refs}

\end{document}